# Towards a general diffusion-based information quality assessment model

A. Lopes Temporao[a,*], M. Temporão[b], C. Vande Kerckhove[c] and F. Abreu Araujo[a,*]

[a]*Institute of Condensed Matter and Nanosciences, Université catholique de Louvain, Louvain-la-Neuve, Belgium*
[b]*Sciences Po Bordeaux, Bordeaux, France, Centre Émile Durkheim, Bordeaux, France*
[c]*Louvain School of Management, Université catholique de Louvain, Louvain-la-Neuve, Belgium*

## Abstract

The rapid and unregulated dissemination of information in the digital era has amplified the global "infodemic," complicating the identification of high quality information. We present a lightweight, interpretable and non-invasive framework for assessing information quality based solely on diffusion dynamics, demonstrated here in the context of academic publications. Using a heterogeneous dataset of 29,264 sciences, technology, engineering, mathematics (STEM) and social science papers from ArnetMiner and OpenAlex, we model the diffusion network of each paper as a set of three theoretically motivated features: diversity, timeliness, and salience. A Generalized Additive Model (GAM) trained on these features achieved Pearson correlations of 0.834 for next-year citation gain and up to 95.62% accuracy in predicting high-impact papers. Feature relevance studies reveal timeliness and salience as the most robust predictors, while diversity offers less stable benefits in the academic setting but may be more informative in social media contexts. The framework's transparency, domain-agnostic design, and minimal feature requirements position it as a scalable tool for global information quality assessment, opening new avenues for moving beyond binary credibility labels toward richer, diffusion-informed evaluation metrics.

[*]Corresponding author. E-mail address: anthony.lopes@uclouvain.be, flavio.abreuaraujo@uclouvain.be

*September 08, 2025*

**Main**

In the era of big data and widespread AI adoption, the quality of information is critical for ensuring the accuracy, fairness, and scalability of machine learning models. Poor data quality can lead to unreliable predictions, bias, and suboptimal decision-making, with consequences that extend far beyond technical performance [1,2]. This challenge has become particularly urgent in the context of a global infodemic: a rapid and unregulated proliferation of false, misleading, or contextually distorted content with far-reaching societal, political, and economic impacts [3,4,5,6,7]. Misinformation has undermined public health responses, eroded trust in democratic institutions, and destabilized economies, yet current detection, prevention, and mitigation strategies remain fragmented and reactive [8]. Addressing this challenge requires a concerted effort to enhance data quality, which would not only improve model performance but also promote more accurate and reliable information dissemination.

Current methodologies for addressing the infodemic predominantly rely on content-based features [9,10,11,12,13], network-based features [14,15,16,17,18], or a combination of both [19,20,21]. Content-based approaches typically analyze linguistic patterns, semantic cues, or source credibility, while network-based methods examine the structure and dynamics of information dissemination across social platforms. Although these strategies offer valuable insights, they face notable limitations. The use of black-box models that hinder interpretability and raise concerns regarding fairness and accountability, particularly in the absence of large-scale, high-quality labeled datasets [22]. These models also suffer from poor generalizability due to the inherently multimodal nature of online information, which increasingly incorporates images, videos, and voice recordings alongside text [23]. Additionally, the evolving nature of information, where content may initially appear factual but later be reclassified, poses significant challenges for static detection models [24]. This temporal volatility is rarely accounted for, leading to potential misclassification. Network-based models, while effective at identifying coordinated behavior or virality patterns, are often unsuitable for early detection, as they require information to have already diffused across the network [25]. Recently, Large Language Models (LLMs) have been explored for automated fact-checking due to their advanced natural language understanding. However, their susceptibility to hallucinations (the generation of factually incorrect yet plausible text) significantly limits their reliability as verification tools [26,27]. This issue is further exacerbated by temporal volatility, as shifting contexts and emerging information increase the likelihood of inaccurate or outdated responses [28].

To address these limitations, this paper proposes a shift from binary misinformation detection toward a broader, more principled framework centered on information quality assessment. Unlike misinformation, which typically focuses on the presence of falsehoods or intent to deceive, information quality captures a different spectrum of attributes, including accuracy, credibility, contextual relevance, consistency, and timeliness [29]. However, much of this work remains user-centric, focusing on how individuals perceive and react to information quality, without accounting for how those perceptions form, evolve, or feed back into future information behaviors [24]. Contextual factors such as task type, website design, and social influence are often studied in isolation, and the dynamic nature of information processing is rarely integrated. This paper addresses these gaps by proposing a generalized, explainable, diffusion-based model of information quality that captures how content spreads, gains credibility, and evolves in collective perception over time. By focusing solely on the diffusion of information, the model mitigates the limitations of content-based approaches such as modality dependence, lack of generalizability, and label scarcity, while remaining domain-agnostic, interpretable, and adaptable to real-time dynamics. To overcome the challenge of defining ground truth, we validate the model using academic citation dynamics, where long-term expert reception serves as a structured proxy for information quality. Citation networks have been extensively studied for citation prediction and impact evaluation [30,31,32,33], making them a well-established and data-rich environment for assessing the temporal and structural indicators of content quality at scale.

Here we show that a simple, interpretable model based entirely on citation diffusion features can predict long-term scientific impact across both science, technology, engineering and mathematics (STEM) and social science domains. Using generalized additive models trained on stratified citation datasets and evaluated across multiple time windows, we find that sustained attention and early endorsement by influential sources are the strongest predictors of future citation gain which is a practical proxy for paper quality in this study. While demonstrated here in the context of scholarly dissemination, the framework is, by essence, domain-agnostic and can be applied to assess the quality of any information that diffuses through a network. These results highlight the potential of diffusion-based signatures as early indicators of information quality and open the door to global, continuous assessment methods that move beyond binary classification of information.

**Results**

This section begins by detailing the construction of the diffusion network and the extraction of diversity, timeliness, and salience features, followed by the design and evaluation of a domain-specific academic quality model and pipeline. We then present global performance metrics, investigate feature contributions through correlation and prediction tasks, and conclude with feature relevance studies that quantify the relative importance and stability of each feature across multiple time windows.

To assess information quality we model its diffusion network as a directed graph formed by its incoming links where each edge representing another actor sharing the information. This network evolves over time, reflecting how the information propagates through a sub-network of connected people. Building on this structure, we define a global model of information quality as the sum of nonlinear functions applied to features extracted from the diffusion network, capturing how different aspects of diffusion relate to higher quality. Specifically, the model is trained on three diffusion features: diversity, timeliness, and salience, which encode intuitive dimensions of information quality we hypothesized: content that spreads across heterogeneous communities (diversity), maintains temporal relevance through sustained sharing (timeliness) [34], and is shared by trustworthy sources (salience) [35] is more likely to be of higher quality.

To evaluate the effectiveness of the proposed diffusion-based model of information quality, we conducted experiments using a heterogeneous dataset of 29,264 academic publications spanning both STEM and social science disciplines, sourced from ArnetMiner [36] and the OpenAlex API, respectively. The ArnetMiner dataset only includes citation data up to the year 2014, so to ensure temporal compatibility, we restricted the OpenAlex sampling to papers published up to 2014 as well. The dataset was stratified based on total citation counts as of 2014 to ensure balanced representation across citation distributions, with approximately equal proportions of STEM and social science papers. This cross-domain design enables the evaluation of our model's generalizability across divergent publication norms and citation dynamics. For all experiments, we applied 10-fold cross validation and kept a 10% validation set to measure the performance of the model.

The adapted academic paper quality model employs a Generalized Additive Model (GAM) trained on the three aforementioned diffusion-derived features: diversity, timeliness, and salience. Diversity is quantified as the number of distinct citation communities the paper reaches within a two-hop citation network. Timeliness is computed as the mean of exponentially weighted citation counts per year, normalized by publication age to emphasize sustained engagement. Salience is modeled as the number of citations normalized by the impact factor of the journal in which the paper was published. GAM was selected for its ability to capture non-linear relationships while preserving interpretability and modularity, allowing for future extensibility with minimal complexity. In this implementation, we deliberately restrict the feature set to demonstrate the feasability of a lightweight and explainable approach.

To evaluate the predictive validity and robustness of our diffusion-based information quality model, we adopted a controlled experimental setup based on fixed time windows. In this design, each paper is allowed to diffuse for a set number of years and the model is trained using only the citation and diffusion data available within that window. This approach ensures uniform exposure time for all papers and mitigates confounding effects such as citation lag [37], thereby enabling fair comparison across papers and timeframes. The model is then evaluated based on its ability to predict future citation gain, which serves as a proxy for scholarly impact. Specifically, we assess near-term performance by predicting citation increases in the year following the time window and long-term utility by identifying high-impact papers defined by being in the 95th percentile of citation gain using only early diffusion signals. A prediction is deemed accurate if the model score for a paper falls in the 90th percentile, with low-impact papers excluded from this evaluation to avoid skewed accuracy estimates.

The model demonstrated strong performance across both tasks. For near-term prediction, it achieved a Pearson correlation of 0.834 and an $R^2$ value of 0.696 between predicted quality scores and actual citation gains in the following year. For long-term impact identification, the model achieved prediction accuracies of 88.32%, 91.24%, 86.13%, and 95.62% when trained on 5, 10, 15, and 20 years of diffusion data respectively based on citationg gains in the 21st year of diffusion. These results affirm the model's ability to predict meaningful scholarly impact based on early-stage information flow, even with limited historical context.

To further understand the contribution of individual features to the model's performance, we conducted a feature relevance anaysis study using 100 independent runs across all combinations of the three core diffusion-based features: Diversity (D), Timeliness (T), and Salience (S). This analysis evaluated how different subsets of features impacted the model's ability to correlate with future citation gains and the high-impact prediction accuracy of the model across the same four diffusion windows of 5, 10, 15, and 20 years. To ensure statistical rigor, we employed pairwise two-sided t-tests to test for significant differences in model performance across feature sets. Additionally, we computed Cohen's d to measure effect size allowing us to quantify practical relevance alongside statistical significance. To account for multiple comparisons, all p-values were adjusted using the Bonferroni correction.

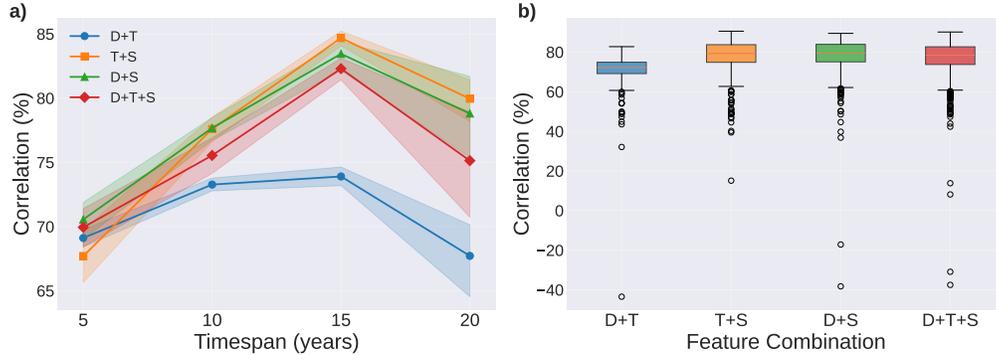

Figure 1: Results of Pearson coefficient of correlation of 100 independent runs for each feature subset and time window. **Plot a)** compares **mean correlation** for different combinations of the three diffusion-based features (Diversity D, Timeliness T, and Salience S), showing the performance of the inclusion of S across all diffusion windows. **Plot b)** shows the **distribution of correlations**, showing the lower variation of the T+S combination.

Correlation analysis (Figure 1) demonstrated that the T+S subset consistently provided the strongest results, peaking at year 15 with a mean correlation of 0.8468 (SD = 0.0296). This significantly outperformed D+T ($p < 0.0001$, Cohen's d = 3.27) and all other subsets. T+S also produced the lowest variance across runs, indicating stable predictive performance. Models including Diversity achieved moderate correlations (e.g., D+S mean = 0.8344 at year 15) but with higher variance, especially at year 20, where outlier runs reduced robustness. The full feature set (D+T+S) yielded strong results overall (mean = 0.8229 at year 15), but did not consistently surpass T+S, suggesting that Diversity introduces redundancy or noise in this academic setting.

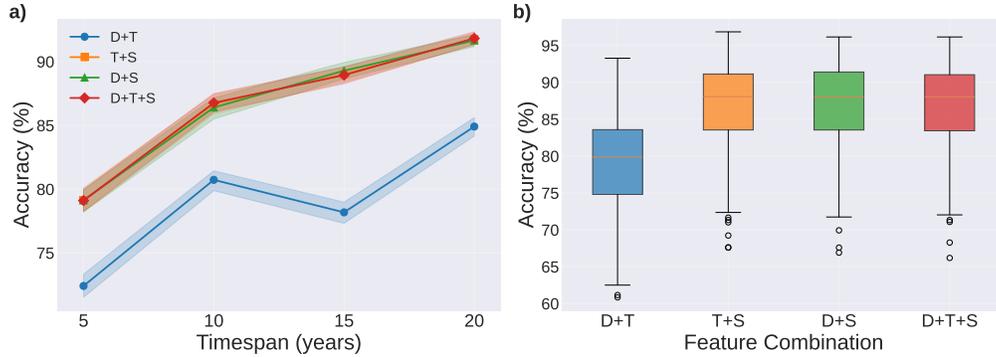

Figure 2: Results of prediction accuracy of 100 independent runs for each feature subset and time window. **Plot a)** compares **mean prediction accuracies** for different combinations of the three diffusion-based features (Diversity D, Timeliness T, and Salience S), showcasing the performance of the (T+S) pair across all diffusion windows. **Plot b)** shows the **distribution of prediction accuracy** scores, emphasizing the robustness of T+S and the relative instability of combinations not involving S.

Prediction accuracy results mirrored the correlation findings (Figure 2). The T+S subset again dominated, reaching a peak of 96.85% accuracy at year 20, and maintaining the highest performance across all windows. By contrast, D+T consistently underperformed (mean = 84.9% at year 20), with large, statistically significant gaps relative to T+S for all timespans ($p < 0.0001$, large effect sizes). Diversity contributed little when added to T+S, as D+T+S did not produce meaningful improvements (mean = 91.8% at year 20). Overall, Salience and Timeliness emerged as the most informative and stable predictors, with Diversity providing marginal or inconsistent benefits.

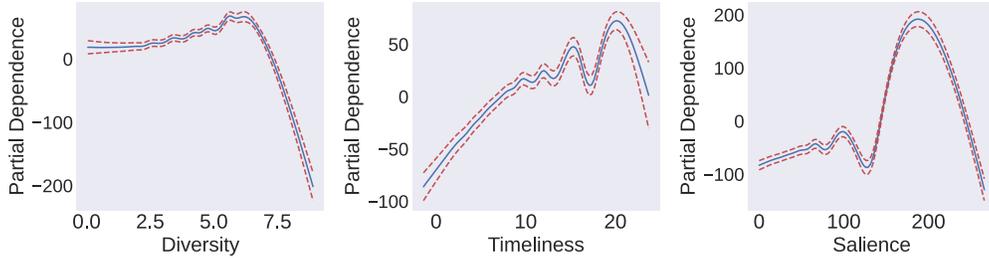

Figure 3: **Partial Dependence Plots** visualizing the marginal effect of each feature on the GAM's output, averaged over all other inputs. Timeliness and salience show strong positive contributions with nonlinear trends, while diversity has a weaker and more volatile effect, with diminishing returns at higher levels of community spread.

These observations were reinforced through partial dependence analysis (cf. Figure 3), which visualizes the marginal effect of each feature on the model's output by averaging over all other inputs. The partial dependence plots (PDPs) illustrate that timeliness has a clear monotonic and positive effect on quality scores, with strong gains at higher levels of sustained citation activity, emphasizing the temporal momentum of diffusion as a key indicator of information value. Salience displays a steep nonlinear contribution that peaks sharply, consistent with the hypothesis that early dissemination by authoritative or high-impact sources strongly boosts perceived quality. In contrast, the contribution of diversity is flatter and more volatile: while modest gains are visible at lower levels, its influence drops steeply beyond a certain threshold, suggesting that diffusion across too many disparate communities may dilute perceived quality.

**Conclusion**

Our findings show that a lightweight, interpretable model grounded in diffusion dynamics can predict future scientific impact across domains. **Timeliness** and **salience** consistently emerged as the strongest predictors of citation gain, underscoring their value as proxies for scholarly quality.

This demonstrates that diffusion patterns can serve as robust indicators of academic paper quality, with potential applications extending beyond scientometrics to broader information ecosystems.

The role of diversity, defined as the number of distinct communities reached, was less stable in the academic setting, offering only marginal gains and sometimes diminishing at higher levels of dispersion. In scholarly networks, excessive spread across disparate communities may dilute coherence, limiting perceived scholarly impact. However, in social media environments such as TikTok, Instagram, Facebook, and X, where users are often trapped in echo chambers and filter bubbles [38,39], diversity could be a key signal. In these contexts, cross-community diffusion may indicate that information is breaking out of insular clusters, making it a potentially valuable feature for assessing the quality and reach of content, including the detection of misinformation.

Despite the empirical performance and theoretical grounding of the model, several limitations must be acknowledged. First, the model is trained exclusively on citation network data, which, while rich and structured, may not generalize to other forms of information diffusion. To truly validate the global applicability of this diffusion-based framework, future work should extend the approach to alternative domains, such as misinformation datasets, Twitter retweet cascades, or any information sharing service where the diffusion is visible, where notions of quality and credibility may follow different social and temporal dynamics.

Second, a core limitation of the proposed model is its dependence on observed diffusion. If an information has not yet diffused or failed to reach visibility, the model is unable to make a quality assessment. This inherently delays early detection of high-quality content and excludes silent or overlooked information that may still be valuable. To address this, future work will explore integrating baseline scores using lightweight, explainable content-based features to support pre-diffusion predictions in a transparent manner.

Third, while GAMs offer a good balance of flexibility and interpretability, they limit the representation of higher-order feature interactions even with the inclusion of interaction terms. Hybrid extensions that preserve modularity but allow interactions (e.g., GAMs with tree-based components [40] or attention over feature groups [41]) may yield further performance gains without compromising transparency.

In summary, our results demonstrate that a diffusion-based, interpretable model can reliably predict academic paper quality, with timeliness and salience proving to be the strongest and most stable indicators of future scholarly impact. While diversity showed limited utility in the academic setting, its potential relevance in more heterogeneous and polarized environments, such as social media, points to broader applications in assessing general information quality and combating misinformation. By focusing on the structural and temporal patterns of how information spreads, this work lays the groundwork for scalable, explainable, and domain-agnostic frameworks that move beyond binary truth labels toward richer, network-informed measures of information value.

## Methods

### Data Acquisition

Our dataset consists of two corpora designed to enable comparative modeling of information diffusion in distinct scientific domains. The first corpus comprises 14,458 STEM papers sourced from the ArnetMiner citation dataset [36]. This dataset includes complete citation diffusion records up to the year 2014. To ensure a balanced representation across impact levels, we employed stratified random sampling based on each paper's cumulative citation count as of 2014. This mitigated skew toward low-citation papers and preserved even distribution across the citation spectrum.

To complement this, we constructed a second corpus of 14,806 social science papers obtained via the OpenAlex API. We applied the same stratified sampling procedure based on 2014 citation counts to ensure comparability with the STEM dataset in both temporal scope and citation profile distribution. The papers were sampled using using the keywords "Medicine", "Psychology", "Sociology" and "Political Science" as a filter for the papers "field" attribute.

For all analyses, we constructed fixed-length time windows of 5, 10, 15, and 20 years, representing the time allowed for each paper to diffuse through the citation network. Each model was trained using only the data available within its respective diffusion window to eliminate potential leakage from future citations and control for time-dependent confounders. Datasets were randomly split into training, validation, and test sets in an 80/10/10 ratio for each window size.

**Global Model and Feature Construction for Academic Paper Quality**

Let us assume that the quality of an information $I$ at time $t$, denoted as $Q_t(I)$ can be modeled as an addition of smoothing functions of specific diffusion features. Let $N$ represent the number of hypotheses $H$ where each hypothesis corresponds to a distinct diffusion feature. Each feature is assigned a smoothing factor $\lambda$, determined during training, such that the overall quality is expressed as:

$$Q_t(I) = \sum_{i=1}^{N} \lambda_i H_{ti}(I) \quad (1)$$

Our academic quality model leverages three main features: **diversity**, **timeliness**, and **salience** designed to capture complementary aspects of early sciientific diffusion. Each feature is computed for every paper using the available citation data within a specified diffusion window. These features quantify structural, temporal, and contextual signals that are potentially predictive of long-term impact. Thus we define the quality of a paper $P$ at time $t$, denoted as $Q_t(P)$, as a sum modeled by three smoothing functions based on these three diffusion features: diversity $D_t(P)$, timeliness $T_t(P)$, and salience $S_{t(P)}$ as shown in the equation below:

$$Q_t(P) = \lambda_1 D_t(P) + \lambda_2 T_t(P) + \lambda_3 S_t(P) + \varepsilon \qquad (2)$$

**Diversity**

The diversity feature $D_t(P)$ captures the structural heterogeneity of a paper's local citation network at time $t$. For each paper $P$, we construct a citation subgraph based on a configurable number of neighborhood hops and apply the fast greedy modularity optimization algorithm to identify community structure [42]. The number of detected communities is used as a proxy for the diversity of scientific discourse surrounding the paper.

$$D_t(P) = \text{Number of communities in local citation graph} \qquad (3)$$

A higher value of $D_t(P)$ suggests broader disciplinary interest. Implementation is based on the `igraph.community_fastgreedy` method, with a `max_depth` parameter controlling subgraph size. This parameter was set to 2 for the model.

**Timeliness**

The timeliness feature $T_t(P)$ quantifies whether the momentum of a paper's citation gain at time $t$. For each paper, we compute yearly citation gains and estimate the trend using the average gradient of the citation trajectory, adjusted for age bias via a penalty term.

$$T_t(P) = \frac{1}{\text{gap}} \sum_{i=0}^{\text{gap}} \nabla g_i - \text{gap} \cdot \text{punish} \qquad (4)$$

Where:

$$\nabla g_i = \text{gained citations}_i - \text{gained citations}_{i-1} \qquad (5)$$

$$\text{gap} = \text{current year} - \text{publishing year} \quad (6)$$

The `punish` hyperparameter calibrates the penalty for late-rising papers. This formulation enables detection of fast-rising papers within the diffusion window. This parameter was left at 1 for the model.

**Salience**

The salience feature $S_t(P)$ contextualizes a paper's impact relative to its venue's historical citation performance at time $t$. It is defined as the difference between the paper's last 2 years citation gained and the current impact factor of the venue:

$$S_t(P) = \text{citations gained last 2 years} - \text{venue impact factor} \quad (7)$$

The venue impact factor is computed from historical data over a configurable time span to match the year from the diffusion window. This helps identify papers that outperform their expected baseline within a given publication context.

**Generalized Additive Model**

The central concept behind Generalized Additive Models (GAMs) [43] is the extension of Generalized Linear Models (GLMs) by replacing the linear combination of predictors with a sum of smooth, data-driven functions. This allows GAMs to model nonlinear relationships while retaining interpretability.

In traditional GLMs, the mean of the response variable $E[Y|x]$ is related to the predictors $x_1, ..., x_p$ via a link function $g$ and a linear predictor:

$$g(E[Y|x]) = \alpha + \beta_1 x_1 + \beta_2 x_2 + ... + \beta_p x_p$$

In contrast, GAMs replace the linear terms with smooth functions $f_j(x_j)$, allowing each predictor to have a flexible, non-parametric effect on the response:

$$g(E[Y|x]) = \alpha + f_1(x_1) + f_2(x_2) + ... + f_p(x_p)$$

These functions $f_j(x_j)$ are learned from the data without imposing a rigid parametric form, enabling the model to capture nonlinear relationships. This flexibility is particularly advantageous when the form of the relationship between a feature and the outcome is unknown or complex.

**Interaction Terms**

While standard GAMs model each feature's effect independently as an additive component, complex phenomena often arise from interactions between features. To capture such dependencies, GAMs can be extended to include interaction terms, typically modeled using tensor product smooths. These allow the model to learn how the combined values of two features jointly influence the response, beyond their individual marginal effects.

Formally, an interaction term between two features $x_i$ and $x_j$ is represented as an additional smooth function $f_{ij}(x_i, x_j)$ in the model:

$$g(E[Y|x]) = \alpha + \underbrace{\sum_j f_j(x_j)}_{\text{Individual Effects}} + \underbrace{\sum_{i<j} f_{ij}(x_i, x_j)}_{\text{Pairwise Interactions}} \qquad (8)$$

These interaction functions are estimated in the same smooth and regularized manner as univariate components, but over a two-dimensional space, allowing the model to learn complex joint patterns.

In our implementation, we incorporate pairwise interaction terms between all three diffusion features to account for interdependencies that may influence information quality in nuanced ways. This model structure en-ables more expressive predictions while preserving interpretability through visualizable two-dimensional partial effects.

**Link Functions**

As with GLMs, GAMs use link functions to relate the expected value of the response variable to the additive predictor. Common examples include:

- Identity link (linear): $g(E[Y|x]) = \mu = \sum f_j(x_j)$

- Log link (Poisson models): $g(E[Y|x]) = \log(\mu) = \sum f_j(x_j)$

- Logit link (binary outcomes): $g(E[Y|x]) = \log\left(\frac{\mu}{1-\mu}\right) = \sum f_j(x_j)$

These link functions are chosen based on the distribution of the response variable, often from the exponential family (Gaussian, Poisson, binomial, etc.). To keep our model simplified we use the identity link as a link function.

**Smooth Functions and Regularization**

The key to GAMs' flexibility lies in the smooth functions $f_j(x_j)$, which are estimated by using splines. These functions are able to capture non-linear features that linear models might miss. However, to avoid overfitting and ensure interpretability, we can impose smoothness penalties during model training.

The smooth functions are often estimated by minimizing a penalized residual sum of squares, as follows:

$$\sum_{i=1}^{n} (y_i - f(x_i))^2 + \lambda \int (f''(t))^2 dt$$

The first term measures the fit to the data, and the second penalizes "wiggliness" of the function $f$, quantified by the integral of its second derivative. The smoothing parameter $\lambda$ governs the trade-off between fidelity to the data and smoothness. A high $\lambda$ enforces smoother functions (potentially linear), while a low $\lambda$ allows for more flexibility.

**Partial Dependence Plots**

To interpret how individual features influence the model's prediction, we employ Partial Dependence Plots (PDPs). A PDP visualizes the marginal effect of a feature on the predicted response in machine learning models such as GAMs. A PDP shows the expected model prediction as a function of one feature, marginalizing over the joint distribution of all other input variables. Formally, for a feature $x_j$, the partial dependence is defined as:

$$PD(x_j) = \left(\frac{1}{n}\right) \sum_{i=1}^{n} \hat{f}(x_j, x_{i,-j})$$

where $\hat{f}$ is the fitted model and $x_{i,-j}$ denotes the values of all features except $x_j$ for the $i$-th observation. This yields a curve showing how predictions change as $x_j$ varies, holding other features fixed.

One key advantage of using GAMs for PDPs is that the additive model structure naturally aligns with the assumption of feature independence made by PDPs. Since each feature's effect is learned independently through smooth functions, the partial effects captured in PDPs closely match the structure of the underlying model.

**Training and Evaluation Pipeline**

We trained GAMs on a stratified dataset composed of 14,458 STEM papers from the ArnetMiner [36] dataset and 14,806 Social Science papers retrieved from the OpenAlex API. All models were trained using only citation diffusion data available within fixed time windows (5, 10, 15, or 20 years post-publication), with a separate model trained for each window length. For each independent run, the training, validation, and test splits (80%/10%/10%) were stratified by citation count and fixed across all diffusion windows to ensure consistent comparisons.

In every run, the GAM was fit to predict citation gain in the year immediately following the diffusion window (e.g., year 11 for a 10-year window). Smoothing parameters ($\lambda$) were independently tuned for each diffusion window using the test set. Model performance was evaluated on the validation set using Pearson correlation for the regression task and high-impact classification accuracy.

For classification, high-impact papers were defined as those in the 95th percentile of citation gain. To mitigate inflated accuracy due to class imbalance, we excluded papers below the 95th percentile from the prediction set, while retaining them during training. Without this exclusion, models achieve artificially high accuracy simply by correctly labeling the majority of low-impact papers, which dominate the dataset under our definition of high-impact papers.

**Feature Relevance Study Tools**

To understand the contribution of different feature types to the model's performance, we conducted a series of feature relevance studies using systematic feature removal. Specifically, we evaluated the model's ability to predict high-impact papers across all possible subsets of our core features: diversity (D), timeliness (T), and salience (S). This resulted in four key combinations: D+S, D+T, T+S, and the full set D+T+S. For each feature subset and for each diffusion time window of 5, 10, 15, and 20 years, we trained 100 models using different random seeds to ensure robustness. We evaluated performance using two primary metrics: (1) prediction accuracy, defined as the proportion of truly high-impact papers also ranked in the 90th percentile of predicted scores, and (2) correlation with next-year citation gain.

To rigorously assess whether observed performance differences across feature sets were statistically meaningful, we conducted **pairwise two-sided t-tests** [44]. These tests evaluate whether the means of two independent sample distributions differ significantly. Given two sets of observed values $X = (x_1, ..., x_n)$ and $Y = (y_1, ..., y_m)$, the t-statistic is computed as:

$$t = \frac{(\bar{x} - \bar{y})}{\sqrt{\frac{s_x^2}{n} + \frac{s_y^2}{m}}} \tag{9}$$

where $\bar{x}$ and $\bar{y}$ are the sample means, $s_x^2$ and $s_y^2$ are the sample variances, and $n$, $m$ are the sample sizes. The resulting p-values determine whether the difference in means is statistically significant.

To complement statistical significance with an estimate of practical impact, we compute **Cohen's d** [45], a standardized effect size defined as:

$$d = \frac{\bar{x} - \bar{y}}{s_p} \tag{10}$$

where $s_p$ is the pooled standard deviation:

$$s_p = \sqrt{\frac{(n-1)s_x^2 + (m-1)s_y^2}{n + m - 2}} \tag{11}$$

Cohen's $d$ helps interpret the magnitude of the observed differences, independent of sample size. As per common conventions, we interpret values of $d \approx 0.2$ as small, $d \approx 0.5$ as medium, and $d \geq 0.8$ as large.

Given that multiple feature combinations were compared across multiple time windows, we applied the **Bonferroni correction** to control for the increased risk of Type I errors due to multiple hypothesis testing. If $m$ pairwise comparisons are made, and the desired significance level is $\alpha$, the adjusted threshold is:

$$\alpha_{\text{corrected}} = \frac{\alpha}{m} \qquad (12)$$

This conservative correction ensures that the family-wise error rate remains bounded and that only genuinely significant results are reported.

**Data availability**

The datasets collected and generated for the current study are available on request. Furthermore the authors are working on making this dataset publicly available.

**Code availability**

The analysis and code prepared for this study are already available on request. Furthermore the code and data will be deposited in a public deposity upon publication.